 \documentclass[final,3p,times]{elsarticle}

\usepackage{slashed}

\usepackage{graphicx}
 \usepackage{epsfig}
\usepackage{color}
\usepackage{amssymb,hyperref}
\usepackage[nodots]{numcompress}

\newcommand{\be}{\begin{equation}}
\newcommand{\ee}{\end{equation}}
\newcommand{\bea}{\begin{eqnarray}}
\newcommand{\eea}{\end{eqnarray}}

\def\beq{\begin{equation}}
\def\eeq{\end{equation}}
\def\bea{\begin{eqnarray}}
\def\eea{\end{eqnarray}}

\begin{document}

\begin{frontmatter}



\begin{flushright}
LA-UR-14-24564
\end{flushright}

\title{A New Spin on Neutrino Quantum Kinetics} 


\author[LANL,ENG]{Vincenzo Cirigliano} 
\author[UCSD,ENG]{George M. Fuller}
\author[UCSD,ENG]{Alexey Vlasenko}

\address[LANL]{Theoretical Division, Los Alamos National Laboratory, Los Alamos, NM 87545, USA}
\address[ENG]{Neutrino Engineering Institute, New Mexico Consortium, Los Alamos, NM 87545, USA}
\address[UCSD]{Department of Physics, University of California, San Diego, La Jolla, CA 92093-0319, USA}

\begin{abstract}

Recent studies have demonstrated that in anisotropic environments 
a coherent spin-flip term arises in the Quantum Kinetic Equations (QKEs) 
which govern the evolution of neutrino flavor and spin in hot and dense media. 
This term  can mediate neutrino-antineutrino transformation 
for Majorana neutrinos and active-sterile transformation for Dirac neutrinos. 
We discuss the physical origin of the coherent spin-flip term and  
provide explicit expressions for the QKEs in  a two-flavor model  with spherical geometry. 
In this context,  we demonstrate that  coherent neutrino spin transformation depends on the absolute  neutrino mass 
and Majorana phases.

\end{abstract}

%\begin{keyword}
%\end{keyword}

\end{frontmatter}

\section{Introduction}

The evolution of  an ensemble of neutrinos in hot and dense media 
is described by an appropriate set of quantum kinetic equations (QKEs), 
accounting for kinetic, flavor, and the often neglected 
spin degrees of 
freedom~\cite{,Sigl:1993fr,Raffelt:1993kx,McKellar:1994uq,Barbieri:1991fj,Enqvist:1990ad,Strack:2005fk,Volpe:2013lr,Vlasenko:2013fja,Zhang:2013lka,
Serreau:2014cfa}.   
QKEs are the essential tool to obtain a complete 
description of  neutrino transport in the early universe, 
core collapse supernovae, and compact object mergers, 
valid before, during, and after the neutrino decoupling epoch (region).  
A self-consistent treatment of neutrino transport is highly relevant because in  
such environments  neutrinos carry a significant fraction of the energy and entropy, and through their flavor- and energy-dependent 
weak interactions play a key role in setting the neutron-to-proton ratio~\cite{Qian93}, 
a critical input for the nucleosynthesis process.

Recent studies~\cite{Vlasenko:2013fja,Serreau:2014cfa}  have demonstrated 
that the QKEs acquire a coherent spin-flip in regions where the spatial 
(anti)neutrino  fluxes are anisotropic  or where there exist anisotropic matter currents. 
Such anisotropy can exist in a core-collapse supernovae or compact object merger environments. 
This spin-flip term  can mediate neutrino-antineutrino transformation 
for Majorana neutrinos and active-sterile transformation for Dirac neutrinos. 
Moreover,  it was shown in Ref.~\cite{Volpe:2013lr}  that a general treatment of neutrino ensembles 
should include  correlations that pair neutrinos and antineutrinos of opposite momenta.  
The coupling to these new densities  to the standard density matrices 
has been worked out explicitly in Ref.~\cite{Serreau:2014cfa}.
In this work  we neglect these terms  as their effect primarily  generates coherence  of opposite-momentum neutrinos 
only for  very long-wavelength  modes, 
with $\lambda_{\rm de Broglie} \sim \lambda_{\rm scale-hight}$, where  
$\lambda_{\rm scale-hight}$ is the length scale characterizing a given astrophysical environment.      
Significant feedback effects  from the long-wavelength modes could  alter the  analysis presented below, 
and this deserves a separate study.

In this letter we further elaborate on the terms of the QKEs describing  coherent neutrino evolution  (i.e. neglecting inelastic collisions). 
The novel aspects of this  work are:

\begin{itemize}

\item We discuss the physical origin of the coherent spin-flip term 
in the framework of a MSW-like effective hamiltonian, 
in analogy   to the  spin-(flavor) oscillations 
induced by neutrino magnetic moments in a  magnetic field. 

\item We  provide explicit expressions for the  coherent   QKEs in a two-flavor model 
with spherical geometry, amenable for a  computational implementation.  
This is the first step towards a realistic exploration of the impact of 
helicity oscillations in astrophysics environments.

\item We point out the dependence of the QKEs 
(through the neutrino-antineutrino conversion term) 
on  the neutrino absolute mass scale and Majorana phases.
We also compare and contrast neutrino-less double beta decay and 
neutrino spin transformation in astrophysical environments as probes of 
these parameters.

\end{itemize}

\section{Spin-mixing term}
\label{sect:spin-mix}

Refs.~\cite{Vlasenko:2013fja,Serreau:2014cfa}  have pointed out that 
in anisotropic environments  the QKEs entail a new term 
that drives  coherent  conversion between different helicity  states (of any flavor). 
An important feature of the new term is that it induces qualitatively different effects  for Dirac and Majorana neutrinos.  
In the Dirac case,  the mixing term  converts active left-handed neutrinos to sterile right-handed states. 
On the other hand,  in the Majorana case  the mixing term  enables conversion of neutrinos into antineutrinos. 
Given the potentially high impact of the spin-flip term,  here we discuss its physical origin in a framework  
that does not involve the intricacies  of non-equilibrium quantum field theory. 
Indeed, as argued below, the basic physics of this term  can be understood in the case of one-flavor Dirac neutrinos 
even at the first-quantized level. 

Physically,   spin oscillations are induced by  the 
axial-vector potential generated by  forward scattering of neutrinos on  the background matter and background neutrinos themselves. 
To illustrate this point, let us  first  consider the evolution of neutrinos 
in  {\it external} chiral four-vector potentials $\Sigma_{L,R}^\mu$ 
(we will give their explicit expressions later on). 
Since  our discussion parallels the analysis of spin-flip transition induced by a neutrino magnetic moment in an external 
magnetic field~\cite{Fujikawa:1980yx,Schechter:1981hw},  we also include in the interaction Lagrangian 
the familiar  magnetic-moment term. 
Suppressing flavor indices ($\mu_\nu$ and $\Sigma_{L,R}$ are matrices in flavor space)  the interaction Lagrangian is given by  
\be
  {\cal L}_{\rm int} =      - \bar{\nu}_L  \slashed{\Sigma}_R \nu_L - \bar{\nu}_R  \slashed{\Sigma}_L  \nu _R  
+ \left( \frac{\mu_\nu}{2}  \ \bar{\nu}_R \sigma_{\mu \nu} F^{\mu \nu} \nu_L    + {\rm h.c.} \right) ~.
\label{eq:Lint1}
\ee
The Majorana case is obtained by  replacing $\nu_R \to \nu_L^c$, $\Sigma_L   \to - \Sigma_R^T$, 
and setting to zero the diagonal elements  $\mu_\nu^{ii}$ (for  Majorana neutrinos $\mu_\nu^{ji} =  -\mu_\nu^{ij}$).
Given this interaction, 
our goal is to obtain an effective hamiltonian in spin(-flavor)  space, with off-diagonal 
components giving the helicity  mixing~\cite{Schechter:1981hw}. 
Since the  essential physics of spin oscillations is already present in the one-flavor case,  to keep the discussion 
as simple as possible we consider the  case of  one-flavor Dirac neutrinos, with real magnetic moment. 
In this case the interaction Lagrangian is:
\be
 {\cal L}_{\rm int} =   \frac{\mu_\nu}{2}  \ \bar{\nu}   \sigma_{\mu \nu} F^{\mu \nu} \nu 
  -  \frac{1}{2} \bar{\nu}   \slashed{\Sigma}_V \nu
-  \frac{1}{2} \bar{\nu}  \slashed{\Sigma}_A \, \gamma_5  \nu~,
\label{eq:Lint2}
\ee
where we have defined the vector and axial-vector potentials as  $\Sigma_{V,A}^\mu  \equiv \Sigma_L^\mu \pm \Sigma_R^\mu = (\Sigma_{V,A}^0, 
\vec{\Sigma}_{V,A})$. 

In a first-quantized approach~\cite{Fujikawa:1980yx}, the  Dirac Hamiltonian corresponding to the  interaction  (\ref{eq:Lint2}) is
\be
H = H_0  + \Delta H~,  \qquad \qquad  H_0 =  \hat{p} \cdot  \vec{\alpha}   + \beta  m~,  \qquad \qquad 
\Delta H =     \mu_\nu  \, \beta  \  \vec{\Sigma} \cdot  \vec{B}   +  \left( \Sigma_V^0 - \vec{\Sigma}_V \cdot \vec{\alpha} \right) 
+  \left( \gamma_5 \Sigma_A^0 - \vec{\Sigma}_A \cdot \vec{\Sigma} \right)~, 
\ee
with  $\beta = \gamma^0$,  $\vec{\alpha} = \gamma^0 \vec{\gamma}$, and $\vec{\Sigma} =  {\rm diag} (\vec{\sigma}, \vec{\sigma})$. 
Defining the helicity operator $h \equiv \hat{p} \cdot \Sigma$, 
already at this level one sees that while  $[H_0, h]=0$, in general  $ [\Delta H, h] \neq 0$, unless   $\vec{\Sigma}_A$ and $\vec{B}$ are 
parallel to the momentum $\vec{p}$.  So the energy  eigenstates  are in general mixtures of helicity eigenstates, and we reach the conclusion that   
magnetic fields {\it and / or} axial-vector potentials transverse to the direction of motion induce helicity oscillations. 

To quantify the helicity mixing effect, it is more convenient to work within the  second-quantized quantum field theory approach~\cite{Schechter:1981hw}.  
One can define the $2 \times 2$  effective hamiltonian in helicity space ${\cal H}_{hh'}$  by computing transition amplitudes 
between  massive neutrino  states    labeled by   momentum $\vec{p}$ and  helicity  $h \in \{ L,R \}$
namely 
\be
\langle  \vec{p}',  h'  \, | \,   \vec{p}, h \rangle   \equiv - i (2\pi)^4  \, 2  E_{\vec{p}}  \,  \delta^{(4)} (p - p') \  {\cal H}_{h'  h} (p).
\label{eq:amplitude}
\ee
To first order in the interaction (\ref{eq:Lint2}) and to all orders in $m/|\vec{p}|$ (with the notation $p \equiv |\vec{p}|$,   $E = \sqrt{m^2 + p^2}$), 
following the steps outlined in  the Appendix  we find
\bea
\label{eq:Heff1}
{\cal H}_{LL} (p)  & = & 
\frac{E + p}{4E}
\Big\{
- 4 r(p)  \, \mu_\nu  \, \hat{p} \cdot \vec{B}  - ( 1 - r(p)^2 ) \Sigma_A^0    + (1 + r(p)^2)  \hat{p} \cdot \vec{\Sigma}_A   + 
( 1 + r(p)^2 ) \Sigma_V^0    - (1 - r(p)^2)  \hat{p} \cdot \vec{\Sigma}_V   
\Big\}
\\
\label{eq:Heff2}
{\cal H}_{RR}  (p) & = & 
\frac{E + p}{4E}
\Big\{
+ 4 r(p) \,   \mu_\nu  \, \hat{p} \cdot \vec{B}  + ( 1 - r(p)^2 ) \Sigma_A^0    - (1 + r(p)^2)  \hat{p} \cdot \vec{\Sigma}_A   + 
( 1 + r(p)^2 ) \Sigma_V^0    - (1 - r(p)^2)  \hat{p} \cdot \vec{\Sigma}_V   
\Big\}
\\
\label{eq:Heff3}
{\cal H}_{LR} (p)  & = & 
\frac{E + p}{2E}
\Big\{ 
(1 + r(p)^2)   \,  \mu_\nu  \, \hat{x}_+ \cdot \vec{B}    \ -  \   r(p) \,   \hat{x}_+  \cdot \vec{\Sigma}_A  
\Big\}
\\
{\cal H}_{RL} (p) & = & 
\frac{E + p}{2E}
\Big\{ 
(1 + r(p)^2)   \,  \mu_\nu  \, \hat{x}_+^* \cdot \vec{B}   \ - \    r(p) \,   \hat{x}_+^* \cdot \vec{\Sigma}_A  
\Big\}~, 
\label{eq:Heff}
\eea
where 
\be
\label{eq:massfactors}
r(p) = \frac{m}{E + p} \qquad \qquad 1 + r(p)^2  =  \frac{2 E}{E + p} \qquad  \qquad 1 - r(p)^2 = \frac{2 p}{E + p}~, 
\ee 
and $\hat{x}_+ \equiv e^{i \phi_p}  (\hat{x}_1    + i \hat{x}_2)$  
with $\hat{x}_{1,2}$ defined so that $(\hat{x}_1, \hat{x}_2, \hat{p})$ form  a right-handed triad.
The choice of  $\hat{x}_{1,2}$ orthogonal to $\hat{p}$ is arbitrary up to a rotation  
along the $\hat{p}$ axis.  We use here  the ``standard  gauge" specified by 
choosing the same azimuthal angle for $\hat{x}_1$ and $\hat{p}$   ($\phi_{{x}_1} = \phi_{{p}}$)~\footnote{In Ref.~\cite{Vlasenko:2013fja} 
a different ``gauge" was used, in which a rotation by $- \phi_{{p}}$ was made in the  $\hat{x}_1$-$\hat{x}_2$ plane.  With this choice 
the phase factors $e^{\pm i  \phi_p}$ disappear from all  formal expressions, 
but the algebra to obtain dot products of $\hat{x}_{1,2} (p)$ with other vectors is more cumbersome.}, 
with unit vectors cartesian coordinates 
expressed in terms of the polar and azimuthal angles $(\theta_p, \phi_p)$ by:
\bea
\label{eq:phat}
\hat{p} &=& (\sin \theta_p \cos \phi_p  \ , \   \sin \theta_p \sin \phi_p \ , \  \cos \theta_p) \\
\label{eq:x1}
\hat{x}_1 &=& (\cos \theta_p \cos \phi_p \ , \  \cos \theta_p \sin \phi_p \ ,  \ -  \sin  \theta_p) \\
\label{eq:x2}
\hat{x}_2 &=& ( -  \sin \phi_p \ , \   \cos  \phi_p \ , \  0)~.
\eea
From the results in (\ref{eq:Heff1}-\ref{eq:Heff}) one sees explicitly that 
helicity mixing occurs only due to components of $\vec{B}$ and $\vec{\Sigma}_A$ transverse to the momentum. 
Note that  factors of $r(p)$ involving one power of mass appear whenever  needed  to provide the 
appropriate helicity flip: in absence of mass,  axial-vector couplings are helicity conserving 
while magnetic dipole couplings are helicity-flipping.    
Note that $r(p)$  provides a suppression factor 
for  axial-induced spin flip amplitude at $|\vec{p}| \gg m$, while it is $O(1)$ at 
$|\vec{p}| \leq m$. 
Besides displaying helicity-flip transitions,  the results in 
(\ref{eq:Heff1}-\ref{eq:Heff})  also encode  the known medium birefringence effect~\cite{Fukugita:1987ag}: 
from parity-violating interactions ($\Sigma_A \neq 0$)  
left-handed and right-handed  states of momentum $\vec{p}$ acquire different energy shifts, 
with energy  splitting proportional  to  $\Sigma_A^0  - \vec{\Sigma}_A \cdot \hat{p}$. 
Finally,  taking the limit  $m/|\vec{p}| \ll 1$,  these  results reproduce the  findings of 
Refs.~\cite{Vlasenko:2013fja,Serreau:2014cfa},  where the more general multi-flavor case was considered.

While so far we have treated the potentials $\Sigma_{V,A}^\mu$ as external fields, 
in a  complete calculation these are induced  by  
forward scattering on a background of matter and (anti)neutrinos. 
The only difference with respect to the standard MSW~\cite{Wolfenstein78,Mikheyev85,Balantekin:2007kx} 
analyses is that here 
we keep non-zero space-like components of the  matter- and neutrino-induced potentials ($\vec{\Sigma}_{V,A} \neq 0$), 
as done for example in Ref.~\cite{Fuller87}.
The explicit expressions (given below) are not crucial to understand the physical origin of  the helicity mixing effect, 
the key point being the spin-dependent  axial coupling.   
So in summary,  neutrino interactions in a non-isotropic  medium   
induce a coupling of the neutrino axial current  to an axial-vector potential  
$\Sigma_A^\mu \equiv  \Sigma_L^\mu - \Sigma_R^\mu = 
 (\Sigma_A^0, \vec{\Sigma}_A)$  (see Eq. (\ref{eq:Lint2})).  
The time-like component $\Sigma_A^0$ induces the well known birefringence effect. 
The space-like potential $\vec{\Sigma}_A$ has a twofold effect: 
(i)  its component  $\vec{\Sigma}_A \cdot \hat p$ parallel to the neutrino propagation gives 
an additional contribution to the energy splitting of $L$ and $R$ states;  
(ii) its component   transverse to $\vec{p}$  induces  mixing of the $L$ and $R$ states.  
In general these effects are flavor dependent, as $\Sigma_A^\mu$ carries flavor indices.

\section{QKE's for  coherent neutrino evolution}
\label{sect:ensemble}

Having established the existence of the helicity-mixing term in the effective hamiltonian 
through simple quantum-mechanical considerations,  we next summarize  how this new term 
appears in the QKEs~\cite{Vlasenko:2013fja,Serreau:2014cfa}.
In Ref.~\cite{Vlasenko:2013fja}  the  QKEs describing the evolution of Majorana neutrinos  
were derived using field-theoretic methods.  These QKEs generalize earlier 
work~\cite{,Sigl:1993fr,Raffelt:1993kx,McKellar:1994uq,Barbieri:1991fj,Enqvist:1990ad,Strack:2005fk} in two respects: 
(i)  They include spin degrees of freedom; 
(ii)  They include effects up to second order in small ratios of scales  
characterizing the neutrino environments we are interested in. 
Specifically,  we treat  neutrino masses,  mass-splitting,  and matter potentials induced by forward scattering,  
as well as  external gradients  as much smaller than 
the typical neutrino energy scale $E$, set by the temperature or chemical potential: namely  $m_\nu/E \sim \Delta m_\nu/E \sim  \Sigma_{\rm forward}/E 
\sim \partial_X/E \sim O(\epsilon)$~\footnote{In the early universe, the small lepton number implies  $\Sigma_{\rm forward}  \sim G_F n_e  \ll m_\nu \sim \Delta m_\nu$. 
This is not the case in supernovae.}.  
The inelastic  scattering can also be characterized by a potential  $\Sigma_{\rm inelastic} \sim \Sigma_{\rm forward} \times G_F  E^2$
which we therefore   power-count as  $\Sigma_{\rm inelastic}/E \sim O(\epsilon^2)$. 
This power-counting  is tantamount to the  statement  that physical quantities 
vary slowly on the scale of the neutrino de Broglie wavelength. 

\subsection{Neutrino density matrices}

QKEs are the evolution equations for suitably defined dynamical quantities 
that characterize a neutrino ensemble, which we will refer to (with slight abuse of language) 
as neutrino density matrices. 
In  the most general  terms a neutrino ensemble is described by 
the set  of all $2n$-field Green's functions, 
encoding  $n$-particle correlations. 
These obey  coupled  integro-differential equations, 
equivalent to the BBGKY equations~\cite{Calzetta:1988qy}.  
As discussed in Refs.~\cite{Sigl:1993fr,Vlasenko:2013fja}, for weakly interacting neutrinos ($\Sigma/E \sim O(\epsilon, \epsilon^2)$)  
the set of coupled equations  can be truncated by  
using perturbation theory to express
all higher order Green's functions 
in terms of the two-point functions.   
In this case the neutrino ensemble is characterized by the full set of one-particle correlations.\footnote{As discussed 
in the introduction, we neglect here 
correlations that pair particles and antiparticles of opposite momenta~\cite{Volpe:2013lr,Serreau:2014cfa}.}
One-particle states of  massive neutrinos and antineutrinos are
specified by the three-momentum $\vec{p}$,  the helicity $h \in \{L,R \}$, and  the  family label $i$ (for eigenstates of mass $m_i$),  
with corresponding  annihilation operators  
$a_{i, \vec{p}, h}$ and  $b_{j, \vec{p}, h}$ 
satisfying the canonical anti-commutation relations 
$ \{ a_{i, \vec{p}, h},
a^{\dagger}_{j, \vec{p}', h'} 
\} = (2 \pi)^3  \, 2 \, \omega_i (\vec{p})  \, \delta_{h h'} \, \delta _{ij} \, \delta^{(3)} (\vec{p} - \vec{p}')$, {\it etc.},  
where $\omega_i (\vec{p}) = \sqrt{\vec{p}^2 + m_i^2}$.
Then, the ensemble is specified by  the matrices  $f_{h h'}^{ij} (\vec{p})$ and 
$\bar{f}_{h h'}^{ij} (\vec{p})$ defined by 
\bea
\langle 
a^\dagger_{j, \vec{p}', h'}  \,  a_{i, \vec{p}, h}
 \rangle   &=&  (2 \pi)^3 \, 2 n_{ij} (\vec{p})  \, \delta^{(3)} (\vec{p} - \vec{p}' ) \  f_{h h'}^{ij} (\vec{p}) ~, \quad 
\\
\langle
b^\dagger_{i, \vec{p}', h'}  \,  b_{j, \vec{p}, h}
\rangle   &=&  (2 \pi)^3 \, 2 n_{ij} (\vec{p})  \, \delta^{(3)} (\vec{p} - \vec{p}' ) \  \bar{f}_{h h'}^{ij} (\vec{p}) ~, 
\eea
where $\langle \dots \rangle$ denotes the ensemble average and 
the normalization factor can be chosen as $n_{ij}    =  2 \omega_{i} \omega_{j}/(\omega_i + \omega_j)$.~\footnote{The interchange $i \leftrightarrow j$ in 
the definition of antiparticle distribution matrices is chosen so that  under unitary transformations   $\nu'  = U \nu$,  $f$ and $\bar{f}$ transform 
in the same way, i.e.  $f ' =      U f  U^\dagger$. } 
For  inhomogeneous backgrounds, the density matrices depend  also on the space-time label, denoted by $x$ in what follows. 

Despite the intimidating index structure,  the physical meaning of the generalized 
density matrices $f_{h h'}^{ij} (\vec{p})$ and 
$ \bar{f}_{h h'}^{ij} (\vec{p})$
is  dictated by simple quantum mechanical considerations: 
the diagonal entries   $f_{hh}^{ii} (\vec{p})$ represent the occupation numbers 
 of  neutrinos of mass $m_i$, momentum $\vec{p}$, and helicity $h$;  
the off diagonal elements    $f_{hh}^{ij} (\vec{p})$  represent  quantum coherence  of states of same helicity and  different mass 
(familiar in the context of neutrino oscillations);  $f_{hh'}^{ii} (\vec{p})$  represent  coherence of  states of different helicity and same mass,  
 and finally   $f_{hh'}^{ij} (\vec{p})$  represent  coherence between states of different helicity and mass.  

In summary,  the basic dynamical object describing  ensembles of neutrinos  and anti-neutrinos 
are the $2 n_f \times 2 n_f$ 
matrices, 
\be
F 
(\vec{p}, x) =
\left(
\begin{array}{cc}
f_{LL} & f_{LR}\\
f_{RL} &  f_{RR}
\end{array}
\right);  
\qquad \qquad \bar{F}
(\vec{p}, x) 
= 
\left(
\begin{array}{cc}
\bar{f}_{RR} & \bar{f}_{RL}\\
\bar{f}_{LR} &  \bar{f}_{LL}
\end{array}
\right), 
\label{eq:FFbar1}
\ee
where we have suppressed the generation indices (each block $f_{h h'}$ is a square $n_f \times n_f$ matrix). 
For  Dirac neutrinos, one needs  both $F$ and $\bar{F}$, with $f_{LL}$ and $\bar{f}_{RR}$ describing  active states.  
For  Majorana neutrinos,  one can choose the phases so that $a_i (\vec{p},h) = b_i (\vec{p}, h)$  and therefore  
$f_{hh'} = \bar{f}_{hh'}^T$ (transposition acts  on flavor indices). Therefore the dynamics is specified by 
$f \equiv f_{LL}$,  $ \bar{f} \equiv \bar{f}_{RR} = f_{RR}^T$,  and $\phi \equiv  f_{LR}$,  and one needs evolution equations only for the matrix ${\cal F}$~\cite{Vlasenko:2013fja}:
\be
F \to {\cal F} = 
\left(
\begin{array}{cc}
f &  \phi \\
\phi^\dagger  &  \bar{f}^T
\end{array}
\right)~.
\ee 
Strictly speaking,  the above discussion in terms of creation and annihilation operators  
makes sense  only within the mass eigenstate basis~\cite{Giunti:1991cb}.  
One can still define ``flavor basis" density matrices  
$f_{\alpha \beta}$ in terms of the mass-basis $f_{ij}$  as 
$f_{\alpha \beta} = U_{\alpha i} f_{ij} U^*_{\beta j}$, 
where $U$ is  the unitary transformation  $\nu_\alpha  = U_{\alpha i}  \nu_i$
that puts the inverse neutrino propagator  in diagonal form. 
While the QKEs 
can be written in any basis,  we give our results below  in the ``flavor"  basis.

\subsection{Anatomy of the QKEs}

A detailed derivation of the QKEs using field-theoretic methods is given in Ref.~\cite{Vlasenko:2013fja}.   
Keeping terms up to $O(\epsilon^2)$ in the power counting discussed earlier on,  
the  QKEs take the compact   $2 n_f \times 2 n_f$ form:
\bea
\label{eq:qkec1}
D_{\vec{p},x}  \   F  (\vec{p},x)    \ = \   -i  \,  \big[  \,   H(\vec{p}, x)   \, , \, F (\vec{p},x)  \,  \big]   \ + \   C (\vec{p}, x) ~; 
\qquad  \qquad 
\bar{D}_{\vec{p},x}  \   \bar{F}  (\vec{p},x)    \ = \  -i  \, \big[  \,    \bar{H} (\vec{p}, x)   \, , \, \bar{F}  (\vec{p},x)  \,  \big]   \ + \   \bar{C} (\vec{p}, x) ~.
\eea
The  differential operator on the left-hand side  generalizes the  usual ``Vlasov" term  of transport equations. 
The first term on the right-hand side controls 
coherent evolution due to mass and forward scattering,  
generalizing  the standard MSW~\cite{Wolfenstein78,Mikheyev85,Balantekin:2007kx}. 
Finally, the second term  on the right-hand side encodes inelastic collisions and generalizes the 
standard Boltzmann collision term used in supernova 
neutrino analyses~\cite{Keil:2003qy,Mezzacappa:2005lr,Kotake:2006fk,Brandt:2011mz,Ellinger:2013gf,
Cherry:2012lu,
Sarikas:2012fk,
Mirizzi:2012qy,
Cherry:2013lr}. 
Here we  focus on the ``Vlasov"-type differential operators ($D$, $\bar D$) and 
the hamiltonian-like operators ($H$, $\bar H$),  describing coherent 
neutrino  evolution.  
The analysis of inelastic collisions ($C$, $\bar C$) was outlined  in Ref.~~\cite{Vlasenko:2013fja}, 
where only a small subset of the contributions to $C$ and $\bar C$ was explicitly calculated. 
Full details on the collision terms will be presented elsewhere~\cite{collision}. 

In order to provide the explicit form of the various operators appearing in (\ref{eq:qkec1}), 
it is extremely useful to introduce the following notation. 
Given an ultra-relativistic neutrino of momentum $\vec{p}$, 
one can naturally  introduce a basis  formed by 
two light-like four-vectors  $n^\mu (p) = (1, \hat{p})$ and $\bar{n}^\mu (p) = (1, -\hat{p})$
(satisfying $n\cdot n = \bar n \cdot \bar n = 0$,  $ n \cdot \bar n = 2$) and 
two transverse four vectors  $x^\mu_{1,2} (p) =  (0,  \hat{x}_{1,2} )$ so  that $n \cdot x_i = \bar n \cdot x_i = 0$ 
and $x_i \cdot x_j = - \delta_{ij}$. 
As discussed below Eq.~(\ref{eq:massfactors}), $\hat p$ and the space-like components $\hat{x}_{1,2}$  of $x^\mu_{1,2}$ 
form a right-handed triad. 

\begin{figure}[t]
\centering
\vspace{-0.5cm}
\includegraphics[width=2.in]{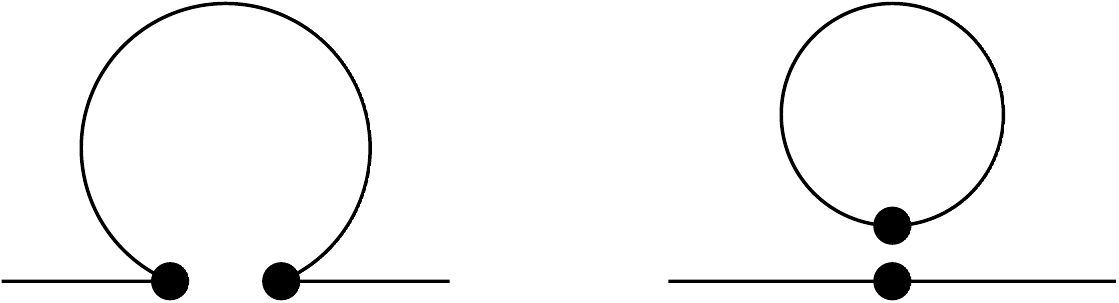}
\caption{Feynman graphs contributing to $\Sigma (x)$. 
External lines represent neutrinos. 
Internal  lines  represent $\nu, e, n, p$  propagators.}
\label{fig:feynman1}
\vspace{-0.5cm}
\end{figure}

\begin{figure}[t]
\vspace{-0.5in}
\centering
\includegraphics[width=4.in]{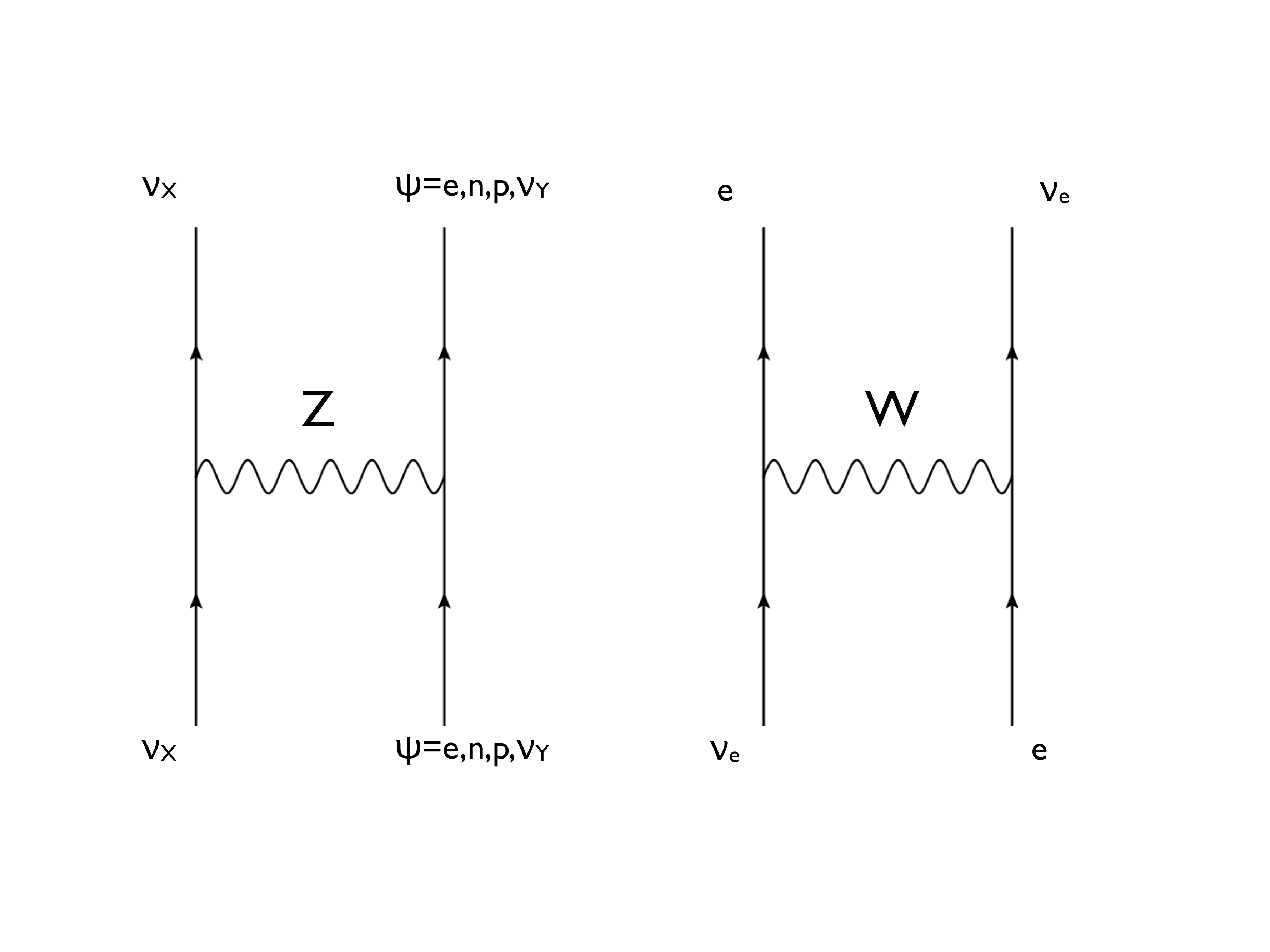}
\vspace{-.5in}
\caption{Tree-level Feynman graphs whose forward-scattering contributions 
generates the 4-vector potential $\Sigma^\mu (x)$. }
\label{fig:msw}
\end{figure}

The key ingredients controlling coherent neutrino evolution 
are the neutrino mass matrix $m$ and the
4-potential induced by forward scattering on matter and other neutrinos. 
In the non-equilibrium field-theory approach,  forward scattering is encoded in the one-loop  
self-energy diagrams of  Fig.~\ref{fig:feynman1}.   In the more familiar  amplitude-based approach this 
physics is described by the diagrams in Fig.~\ref{fig:msw}. 
The chiral 4-potentials  can be arranged in the $2 n_f \times 2 n_f$ structure 
\be
\Sigma^\mu  (x)  = 
\left(
\begin{array}{cc}
\Sigma_R^\mu  (x)  & 0 \\
0 &  \Sigma_L^\mu  (x)
\end{array}
\right)~.
\label{eq:sigmafull}
\ee
$\Sigma_R$  and  $\Sigma_L$ are the potentials for left-handed and right-handed neutrinos, respectively.  
For Dirac neutrinos $\Sigma_R \neq 0$ while $\Sigma_L \propto G_F  m^2 \sim O (\epsilon^3) $ (massless right-handed neutrinos do not interact). 
On the other hand, in the Majorana case  one has $\Sigma_L = - \Sigma_R^T$, with transposition acting on flavor indices. 
The potential induced by a  background of electrons and positrons is given  for any geometry by the following 
expressions:   
\bea
\left[ \Sigma_R^\mu  \ \Big \vert_{e}  
\right]_{IJ}
&=& 
2\sqrt{2} G_F   \, 
\left[  \left(  \delta_{eI} \delta_{eJ}  + \ \delta_{IJ}  \, \left(\sin^2\theta_W-\frac{1}{2}\right)\right)J_{(e_L)}^\mu+ 
\delta_{IJ} \ \sin^2\theta_W \,  J_{(e_R)}^\mu \right]
\\
J^\mu_{(e_L)}   (x) &=& \int \frac{d ^3q}{(2 \pi)^3}   \  v_{(e)}^\mu (q) \  \Big( f_{e_L} (\vec{q},x) - \bar{f}_{e_R} (\vec{q},x) \Big)~,
\\
J^\mu_{(e_R)}   (x) &=& \int \frac{d ^3q}{(2 \pi)^3}   \  v_{(e)}^\mu (q) \  \Big( f_{e_R} (\vec{q},x) - \bar{f}_{e_L} (\vec{q},x) \Big)~,
\eea
where $v_{(e)}^\mu =    (1, \vec{q} / \sqrt{m_e^2 + q^2} \,  )$,   and we use the notation $f_{e_L} (\vec{q},x)$  ($\bar{f}_{e_L} (\vec{q},x)$)
for  the distribution function of  L-handed electrons (positrons), etc. 
The nucleon-induced potentials have similar expressions , with appropriate 
replacements of the L- and R-handed couplings to the $Z$  and the 
distribution functions $f_{e_L} \to f_{N_L}$, etc. 
For unpolarized electron and nucleon backgrounds of course one has $f_{e_L} = f_{e_R} = (1/2) f_e$, etc., 
and the nucleon contribution to the potential is:
\be
\left[ \Sigma_R^\mu  \ \Big \vert_{N}  
\right]_{IJ}
=  \sqrt{2} G_F   \,  C_V^{(N)}  \   J_{(N)}^\mu   \  \delta_{IJ}~,   \qquad \qquad C_V^{(n)} = - \frac{1}{2}~,  \quad  C_V^{(p)} = \frac{1}{2}  -2  \sin^2 \theta_W~. 
\ee
On the other hand, the neutrino-induced potentials are given by
\bea
\label{eq:Sigma}
\left[ \Sigma_R^\mu  \ \Big \vert_{\nu} \right]_{IJ}  &=& 
 \sqrt{2} G_F 
 \left(
\left[ J^{\mu}_{(\nu)}  \right]_{IJ}  \ + \   \delta_{IJ}   \  {\rm Tr}   J^\mu_{(\nu)} 
 \right)
\\
J^\mu_{(\nu)}  (x) &=&    \int \frac{d ^3q}{(2 \pi)^3}   n^\mu (q) \  \Big( f_{LL} (\vec{q},x) - \bar{f}_{RR} (\vec{q},x) \Big)~,
\label{eq:Sigmav2}
\eea
with $n^\mu (q) = (1, \hat q)$. 
For a test-neutrino of three-momentum $\vec{p}$, these potentials  can be  further projected along the basis vectors: 
with light-like component $\Sigma^\kappa \equiv  n (p) \cdot \Sigma$ 
along the neutrino trajectory (in the massless limit); 
and space-like component $\Sigma^i \equiv x^i (p)  \cdot \Sigma$,  transverse to the neutrino trajectory.  
In particular, for the neutrino-induced contribution we find 
 $\Sigma^\kappa (x) \propto    \int d^3 q  \ (1 - \cos \theta_{p q} )  \cdot ( f_{LL} (\vec{q},x) - \bar{f}_{RR} (\vec{q},x))   $, consistently with the familiar results in the literature 
(\cite{Duan:2010fr} and references therein).

In terms of the mass matrix $m$ and the potentials $\Sigma^\mu_{L,R}$, 
the Hamiltonian-like operators controlling the coherent evolution are given by 
\be
H  =
\left(
\begin{array}{cc}
H_R &  H_{LR}  \\
H_{LR}^\dagger   & H_L
\end{array}
\right)~ \qquad 
\bar H  =
\left(
\begin{array}{cc}
\bar{H}_R &  H_{LR}  \\
H_{LR}^\dagger   & \bar{H}_L
\end{array}
\right) ~, 
\ee
with 
\bea
H_R  &=&  \Sigma_R^\kappa   + \frac{1}{2 |\vec{p}|}   \left( m^\dagger m  - \epsilon^{ij} \partial^i \Sigma_R^j   + 4 \Sigma_R^+ \Sigma_R^-\right)
\label{eq:HR}
 \\
H_L  &=&  \Sigma_L^\kappa   + \frac{1}{2 |\vec{p}|}   \left( m  m^\dagger  + \epsilon^{ij} \partial^i \Sigma_L^j   + 4 \Sigma_L^- \Sigma_L^+\right)
\label{eq:HL}
\\
H_{LR}  &=&   - \frac{1}{|\vec{p}|}  \left( \Sigma_R^+  \, m^\dagger - m^\dagger \, \Sigma_L^+ \right) ~, 
\label{eq:Hm}
\eea
where $\Sigma_{L,R}^\pm \equiv (1/2) \,  e^{\pm i \phi} \,   ( x_1 \pm i x_2)_\mu  \,   \Sigma^\mu_{L,R}$.
The antineutrino operators $\bar{H}_{L,R}$ can be obtained from $H_{L,R}$ by flipping the sign of the entire term multiplying $1/(2 |\vec{p}|)$.  
The first two terms in $H_{L,R}$ are included in all analyses of neutrino oscillations in medium. 
$\Sigma^\kappa_{L,R}$ include the usual forward scattering off matter and neutrinos, and are functions of $F, \bar F$ 
thereby introducing  non-linear effects  in the coherent evolution. 
Theo $m^\dagger m/|\vec{p}|$ term encodes vacuum oscillations. 
The additional terms in $H_{L,R}$ and the spin-flip term $H_{LR}$ 
complete the set $O(\epsilon^2)$ terms,  and can be as important as $m^2/|\vec{p}|$ in supernova environments. 
The  spin-flip  term $H_{LR}$ is given in compact matrix form in (\ref{eq:Hm}), and its physical origin 
has been discussed in Section~\ref{sect:spin-mix}.
Note that the spin-flip term $H_{LR}$ depends linearly on the mass matrix $m$,  while 
the vacuum hamiltonian depends on $m^\dagger m$.   Therefore,  as we will show explicitly later, 
the spin-flip term is sensitive to the absolute mass scale of the neutrino spectrum and (for Majorana neutrinos) 
to the Majorana phases.

Finally,    using the compact notation  $\partial^\kappa \equiv  n (p) \cdot  \partial =\partial_t  + \hat{p} \cdot \partial_{\vec{x}}$,  
$ \, \partial^i \equiv x^i  (p) \cdot \partial = \hat{x}_i  \cdot \partial_{\vec{x}}$, 
the generalized Vlasov operators are  (recall that the $2n_f \times 2 n_f$ potential $\Sigma^\mu$ is defined in (\ref{eq:sigmafull}))
\bea
D_{\vec p, x} F (\vec p,x) &=&  \partial^\kappa F  + \frac{1}{2 |\vec{p}|}  \left\{ \Sigma^i , \partial^i F \right\}  - \frac{1}{2}  \left\{  \frac{\partial \Sigma^\kappa}{\partial \vec{x}}, 
\frac{\partial F}{\partial \vec{p}}       \right\}    
\\
\bar{D}_{\vec p,x}  \bar  F  (\vec{p},x) &=&  \partial^\kappa \bar F  -  \frac{1}{2 |\vec{p}|}  \left\{ \Sigma^i , \partial^i \bar{F} \right\}  + \frac{1}{2}  \left\{  \frac{\partial \Sigma^\kappa}{\partial \vec{x}},  \frac{\partial \bar F}{\partial \vec{p}}       \right\}.  \ \ 
\eea
The physical meaning of  $D$ and $\bar D$  becomes more transparent by 
noting that they can be re-written as 
\be
\partial_t +  \frac{1}{2} \{ \partial_{\vec{p}}  \omega_\pm ,    \partial_{\vec{x}}  \  \}  - \frac{1}{2}  \{  \partial_{\vec{x}} \omega_\pm  , \partial_{\vec{p}} \   \}, 
\ee
with 
$\omega_+ =   |\vec{p}|  +  \Sigma^\kappa$ 
 in D and $\omega_- =  |\vec{p}|  -  \Sigma^\kappa $ in $\bar D$.  
Recalling  that   $\omega_{\pm}  (\vec{p})   =  |\vec{p}|  \pm \Sigma^\kappa$ 
are the $O(\epsilon)$    neuutrino ($+$) and anti-neutrino ($-$) hamiltonian operators, 
one sees that  $D$ and $\bar D$  
generalize the total time-derivative operator
$d_t =  \partial_t   + \dot{\vec{x}}  \ \partial_{\vec{x}} +   \dot{\vec{p}}   \ \partial_{\vec{p}} $, 
with 
$\dot{\vec{p}}  = - \partial_{\vec{x}} \, \omega$
 and   
$\dot{\vec{x}}  =  \partial_{\vec{p}} \, \omega$, 
thus encoding 
the familiar
drift and force terms. 

\section{The bulb model}

\begin{figure}[t]
\centering
\includegraphics[width=1.5in]{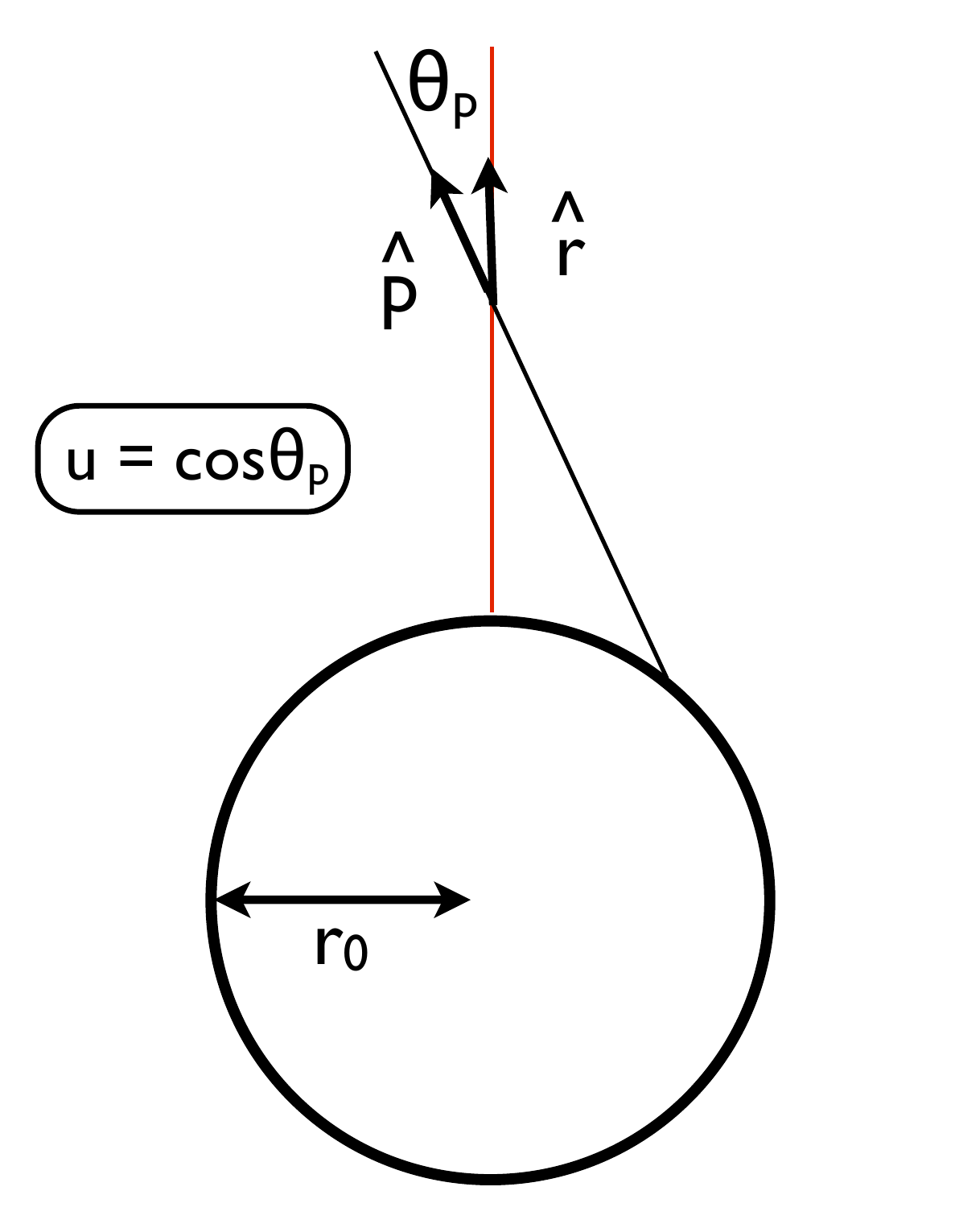}
\caption{Bulb model geometry.}
\label{fig:bulb}
\vspace{-0.1in}
\end{figure}

The simplest realistic setup to explore the impact of  helicity oscillations  in supernovae is 
provided by the so-called bulb model.   
In this model one assumes  spherical symmetry  and further 
assumes that  neutrinos are emitted isotropically, with a given spectrum and luminosity, 
from a sharply defined neutrino-sphere of  radius $r_0$ (see Fig.~\ref{fig:bulb}). 
In this section we provide  explicit expressions for the coherent QKEs   describing 
two-flavor Majorana neutrinos in the bulb model.  
These expressions are amenable to computational implementation and allow us to   
explicitly point out the dependence of QKEs on the absolute neutrino mass scale 
and the  Majorana phase characterizing the two-flavor problem. 
Throughout the discussion,  it is useful to keep in mind that the $\Sigma_{R}^\mu$ potential in this 
geometry only has   time-like  and  radial space-like components (the transverse 
components cancel). 

To make contact with the existing literature on the subject~\cite{Sigl:1993fr,Serreau:2014cfa,deGouvea:2012hg,deGouvea:2013zp},  
we introduce the following notation for the $2 \times 2$ 
blocks of the Majorana density matrix
\be
f =
\left(\begin{array}{cc}
\rho_{e e} & \rho_{e {x}} \\
\rho_{e x}^*  & \rho_{xx}
\end{array}\right)
\qquad \qquad 
\bar f =
\left(\begin{array}{cc}
\rho_{\bar e \bar e} & \rho_{\bar e  \bar{x}} \\
\rho_{\bar e  \bar x}^*  & \rho_{\bar x \bar x}
\end{array}\right)
\qquad \qquad 
\phi=
\left(\begin{array}{cc}
\rho_{e\bar{e}} & \rho_{e\bar{x}} \\
\rho_{x\bar{e}} & \rho_{x\bar{x}}
\end{array}\right)~, 
\ee
so that   the flavor $\times$ spin density matrix  for two flavors ($\nu_e$, $\nu_x$) is given by
\begin{eqnarray}
{\cal F}=\left(\begin{array}{cc}f & \phi \\ \phi^\dagger & \bar{f}^T\end{array}\right)=
\left(\begin{array}{cccc}
\rho_{ee} & \rho_{ex} & \rho_{e\bar{e}} & \rho_{e\bar{x}}\\
\rho_{ex}^\star & \rho_{xx} & \rho_{x\bar{e}} & \rho_{x\bar{x}}\\
\rho_{e\bar{e}}^\star & \rho_{x\bar{e}}^\star & \rho_{\bar{e}\bar{e}} & \rho_{\bar{e}\bar{x}}^\star\\
\rho_{e\bar{x}}^\star & \rho_{x\bar{x}}^\star & \rho_{\bar{e}\bar{x}} & \rho_{\bar{x}\bar{x}}
\end{array}\right)~.
\end{eqnarray}
The coherent QKEs  for the density matrix are
\begin{eqnarray}
D_{\vec{p},x}   \, {\cal F}  (\vec{p}, x) =-i\left[{\cal H}, {\cal F}\right]~, 
\end{eqnarray}
and the assumption of spherical symmetry implies that the density matrix depends only on $p \equiv | \vec{p} |  \simeq E$,  $u \equiv \cos \theta_{p}$,  and $r$, 
so that $  {\cal F}  (\vec{p}, x)  \to  {\cal F} (E, u, r)$.

Both the Vlasov differential operator $D_{\vec{p},x}$ and the Hamiltonian ${\cal H}$ depend  on the 4-potential 
$\Sigma^\mu$  (\ref{eq:sigmafull}),  that in spherical symmetry has a time-like and a space-like radial component
\be
\Sigma^\mu =  
\left(
\Sigma^0_{\rm matter} + \Sigma^0_{\nu}, \   \left(\Sigma^r_{\rm matter} + \Sigma^r_{\nu}  \right)  \,  {\hat{r}}
\right)~.
\ee
Assuming the absence of muons, and neglecting corrections of order $G_F^2$,  
the matter contribution is given by:
\be
\Sigma^0_{\rm matter}=\frac{G_F n_B}{\sqrt{2}} \
\left(\begin{array}{cccc}
3Y_e-1 & 0 & 0 & 0 \\
0 & Y_e-1 & 0 & 0 \\
0 & 0 & -\left(3Y_e-1\right) & 0 \\
0 & 0 & 0 & -\left(Y_e-1\right)
\end{array}\right)  \qquad \qquad 
\Sigma^r_{\rm matter} =  V_{\rm out} \ \Sigma^0_{\rm matter}~, 
\ee
where, $n_B$ is the baryon number density, $Y_e = \left(n_e-n_{\bar{e}}\right)/n_B$ is the electron lepton number fraction, 
and $V_{\rm out}$ is the (radial)  matter outflow speed as a fraction of the speed of light.
Similarly, the neutrino-induced potential is
\be
\Sigma^{0}_\nu  =\sqrt{2}G_F
\left(\begin{array}{cccc}
2J^{0}_{ee}+J^{0}_{xx} & J^0_{ex} & 0 & 0 \\
J^{0\star}_{ex} & 2J^0_{xx}+J^0_{ee} & 0 & 0 \\
0 & 0 & -2J^0_{ee}-J^0_{xx} & -J^{0\star}_{ex} \\
0 & 0 & -J^0_{ex} & -2J^0_{xx}-J^0_{ee}
\end{array}\right) \qquad \qquad \Sigma^r_\nu =  \Sigma^0_\nu  \Big\vert_{J^0_{IJ} \to J^r_{IJ}}~,
\ee
with time-like and radial  components of the neutrino current $J^0_{IJ}$ and $J^r_{IJ}$ given by:
\footnote{
In general the radial component of the current receives an additional contribution~\cite{Serreau:2014cfa} 
$$
\delta J^r_{IJ} = 
 \int_0^{2 \pi} \frac{d\phi'}{2 \pi} 
\int \frac{E' dE'}{2\left(2\pi\right)^2}\int_{u_{\rm min}}^1 du' \sqrt{1-u'^2 
\  }\left[m^\dagger\left(\phi^\dagger  -   \phi^\star\right)   
e^{i \phi'} 
+
e^{- i \phi'}  
\left(\phi   -     \phi^T\right)m\right]_{IJ} ~.
%\to 0~.
$$
This term,  while suppressed by  $m/E$, 
might be important near neutrino-antineutrino resonance. 
In  spherical symmetry, however,  $\delta J^r_{IJ} =0$  as the dynamical 
functions $\phi (E', u', r)$ do not depend on the azimuthal angle  $\phi'$.}
\begin{eqnarray}
J^0_{IJ}&=&\int\frac{E'^2dE'}{\left(2\pi\right)^2}\int_{u_{\rm min}}^1du'
\left[ f_{IJ} \left(E',u'\right)-  \bar{f}_{JI}  \left(E',u'\right)\right]
\label{eq:J0}
\\
J^r_{IJ}&=&\int\frac{E'^2dE'}{\left(2\pi\right)^2}\int_{u_{\rm min}}^1du'\ u' 
\left[ f_{IJ} \left(E',u'\right)-  \bar{f}_{JI}  \left(E',u'\right)\right]~.
%\   +  \  \delta J^r_{IJ}~.
\label{eq:Jr}
\end{eqnarray}

In terms of the potentials explicitly given above,  the Vlasov operator is 
\be
D_{\vec{p},x}   \, {\cal F}  =    u  \ \frac{\partial  {\cal F} }{\partial r} 
+  \frac{1-u^2}{r}   \frac{\partial  {\cal F} }{\partial u} 
-  \frac{1 - u^2}{2  E}  \, \left\{    \Sigma^r  \,  ,  \, \frac{\partial {\cal F}}{\partial r}  \right\}
- \frac{1}{2}  \, \left\{
\frac{\partial \Sigma^0}{\partial r}  - u \frac{\partial \Sigma^r}{\partial r}  \, , \, 
u \frac{\partial  \cal{F} }{\partial E}  + \frac{1- u^2}{E} \  \frac{\partial \cal{F}}{\partial u}  
\right\}~.
\ee
Within  current ``multi-angle"  simulations of the bulb model the terms proportional 
to $\Sigma^r$ and $\Sigma^0$ are usually dropped, retaining only the first two terms of the 
above Vlasov operator. In a consistent analysis to second order in gradients and interactions 
all terms in the above expression should be kept. 

The Hamiltonian contains vacuum, matter and neutrino contributions. 
In the presence of spacelike currents, the matter and neutrino contributions give a spin flip term.  
We break up the terms in the Hamiltonian as follows:
\begin{eqnarray}
{\cal H} = {\cal H}_{\rm vac}+{\cal H}_{\rm matter}+{\cal H}_{\rm \nu}+{\cal H}_{{\rm sf}}~.
\end{eqnarray}
The traceless part of the vacuum Hamiltonian is ($\Delta m^2 \equiv m_2^2 - m_1^2 >0$) 
\begin{eqnarray}
{\cal H}_{\rm vac} = \frac{\Delta m^2}{4 E}
\left(\begin{array}{cccc}
-c_{2\theta} & s_{2\theta} & 0 & 0\\
s_{2\theta} & c_{2\theta} & 0 & 0\\
0 & 0 & -c_{2\theta} & s_{2\theta}\\
0 & 0 & s_{2\theta} & c_{2\theta}
\end{array}\right)~, 
\end{eqnarray}
where $s_{2 \theta} =  \sin 2\theta$,  $c_{2 \theta} = \cos 2 \theta$ and $\theta$ is the two-flavor mixing angle. 
The matter and neutrino  Hamiltonians are
\bea
H_{\rm matter} &=& \Sigma^0_{\rm matter}  \ - \ u  \, \Sigma^r_{\rm matter} 
\\
H_{\nu} &=& \Sigma^0_{\nu} \  -  \ u \, \Sigma^r_{\nu}~.
\eea
The spin-flip Hamiltonian includes a matter term (in the presence of bulk motion of matter) and a neutrino term. 
In $2\times 2$ block form, the-spin flip Hamiltonian is
\begin{eqnarray}
{\cal H}_{\rm sf}=
\left(\begin{array}{cc}
0 &   e^{i \phi_p} 
\left[  \left(H_{\rm sf}^{\rm matter}+H_{\rm sf}^\nu\right)\frac{m^\star}{E}+
\frac{m^\star}{E}\left(H_{\rm sf}^{\rm matter}+H_{\rm sf}^\nu\right)^T 
\right]  \\
e^{- i \phi_p} 
\left[  \left(H_{\rm sf}^{\rm matter}+H_{\rm sf}^\nu\right)\frac{m^\star}{E}+\frac{m^\star}{E}\left(H_{\rm sf}^{\rm matter}+H_{\rm sf}^\nu\right)^T\right]^\dagger & 0
\end{array}\right)~,
\end{eqnarray}
where the phase factor can be set to $e^{i \phi_p} = 1$ in spherical symmetry   
and   the $2\times 2$ matrices $H_{\rm sf}^{\rm matter}$ and $H_{\rm sf}^{\nu}$ are
\begin{eqnarray}
H_{\rm sf}^{\rm matter}=-\frac{G_F n_B}{2\sqrt{2}}V_{\rm out}\sqrt{1-u^2}
\left(\begin{array}{cc}3Y_e-1 & 0 \\ 0 & Y_e-1\end{array}\right)
\end{eqnarray}
\begin{eqnarray}
H_{\rm sf}^{\nu}=-\sqrt{2}G_F\sqrt{1-u^2}\left(\begin{array}{cc}2J^r_{ee}+J^r_{xx} & J^r_{ex} \\ J^{r\star}_{ex} & 2J^r_{xx}+J^r_{ee}\end{array}\right)~.
\end{eqnarray}

As observed in the previous sections, the spin-mixing Hamiltonian depends linearly on the 
neutrino mass matrix. 
In the two-flavor case, the Majorana mass  matrix can be written as:
\be
 m = U^* \, m_d  \, U^\dagger~;  \
\qquad {\rm with} \qquad
m_d = \left( \begin{array}{cc}
m_1 & 0 \\
0  &  m_2
\end{array}
\right) ~, 
\qquad 
U =
\left(
\begin{array}{cc}
c_\theta & s_\theta  \\
- s_\theta & c_\theta
\end{array}
\right)
\times
\left(
\begin{array}{cc}
1 & 0 \\
0  & e^{i \alpha/2}
\end{array}
\right) ~.
\ee
In terms of the observable parameters $\Delta m^2  \equiv m_2^2 - m_1^2 >0$ and $m_0 \equiv   (1/2 )  (m_1 + m_2)$
(so that  $m_{1,2} = m_0  \mp  \Delta m^2/(4 m_0)$) we find:
\begin{equation}
m =
m_0
\left(
\begin{array}{ccc}
c_\theta^2  + e^{-i \alpha} s_\theta^2  &  &  (e^{-i \alpha} - 1)  s_\theta
c_\theta  \\
& & \\
  (e^{-i \alpha} - 1)  s_\theta c_\theta   &  & s_\theta^2 + e^{-i \alpha}
c_\theta^2
\end{array}
\right)
+ \frac{\Delta m^2}{4 m_0}
\left(
\begin{array}{ccc}
- (c_\theta^2  - e^{-i \alpha} s_\theta^2)  &   &  (e^{-i \alpha} + 1)
s_\theta c_\theta  \\
& & \\
  (e^{-i \alpha} + 1)  s_\theta c_\theta   &   & c_\theta^2 e^{-i \alpha}
-  s_\theta^2
\end{array}
 \right) ~.
 \label{eq:majoranamass}
\end{equation}
Eq.~(\ref{eq:majoranamass}) shows explicitly the dependence of the mass matrix on 
the absolute mass scale of the neutrino mass spectrum, $m_0$,  the mixing angle $\theta$, and the 
Majorana phase $\alpha$.  
The phase  $\alpha$ can significantly alter the spin-flavor
mixing structure compared to the Dirac case ($\alpha=0$), for any value of $m_0/ \sqrt{\Delta m^2}$.
For example, in the degenerate limit $m_0 \gg \sqrt{\Delta m^2}$
(in which we expect  ${\cal H}_{\rm sf}$  to have the largest impact) the first term in 
(\ref{eq:majoranamass}) dominates, and a non-zero 
Majorana phase $\alpha$  can induce $O(1)$ off-diagonal terms compared to the vanishing ones in the Dirac case ($\alpha=0$).

In absence of a full-fledged calculation no conclusion can be drawn on the impact 
of Majorana phases on supernova neutrinos. 
However, we find it  very interesting  that at least in principle astrophysical processes 
are sensitive to these parameters.
In fact,  until now neutrino-less double beta decay  experiments  offer the only way to probe 
a subset of these quantities~\cite{deGouvea:2013zba}, 
namely the element $m_{ee}$ of the mass matrix given in  (\ref{eq:majoranamass}).
The detection of large scale neutrino-antineutrino transformation in a supernova neutrino burst 
could provide information complementary to that obtained from neutrino-less double beta  decay searches.

\section{Conclusions}

In this letter we have discussed 
he physical origin of the coherent spin-flip term in the neutrino QKEs 
in the framework of a MSW-like effective hamiltonian.  
The key point is that in anisotropic environments, neutrino forward scattering on matter and other neutrinos 
induces  not only a time-like but also a space-like axial-vector  potential:
the latter couples to neutrino spin and generates helicity mixing. 
We  have also  provided explicit expressions for the  coherent   QKEs in a two-flavor model 
with spherical geometry: this exercise is a necessary 
step towards a computational analysis of the QKEs in astrophysical environments, 
and gives us the opportunity to illustrate 
an under-appreciated point:   through the spin-flip term, 
neutrino evolution is sensitive to the absolute scale of the neutrino  mass spectrum 
and to the Majorana phases. 

While the spin-mixing effect is in general small,  $O(G_F  \times m_\nu / E)$,  
it may become dominant if a resonance occurs.   
In fact,  an exploration within a simplified setup~\cite{Vlasenko:2014bva}  indicates that 
non-linearities can keep the system near resonance thus leading to 
large scale neutrino-antineutrino conversion.
In a different context,  
it has been shown~\cite{deGouvea:2012hg,deGouvea:2013zp} that 
magnatic-induced spin-flavor oscillations can significantly impact 
supernova neutrinos. In fact, 
assuming typical magnetic fields in a supernova envelope ($B \sim  10^{10-12} G$) and 
Majorana transition magnetic moments a factor of 100 larger than the SM values, 
so that $\mu_\nu^{ij} B (r) \sim 10^{-18}  \, {\rm eV} \, (50 {\rm km}/r)^2$,  
Refs~\cite{deGouvea:2012hg,deGouvea:2013zp}  find that  magnetic spin-flip transitions 
lead to significant effects  on collective neutrino oscillations in supernovae.  
A naive estimate based on $\Sigma \sim \sqrt{2} G_F  (n_\nu - n_{\bar{\nu}})$, 
with net neutrino density $n_\nu - n_{\bar{\nu}}$,  suggests that $H_{LR} \geq  10^{-18}$~eV
at $r  \leq 100$~km (for $m_\nu = 0.01$~eV and $|\vec{p}|  = 10$~MeV), in the same ballpark as the  magnetic term. 
While these estimates are rough since they ignore the flavor structure and the 
effects of geometry,    combined  with  the results of Refs.~\cite{deGouvea:2012hg,deGouvea:2013zp}, 
they nevertheless suggest potential implications for supernova neutrinos. 

The conditions for significant neutrino-antineutrino conversion 
require  large $\nu$ luminosities and the presence of  level-crossings (resonances)~\cite{Vlasenko:2014bva}.
These conditions are likely to be realized during the supernova
neutronization burst, where there is a large dominance of $\nu_e$ over the $\bar{\nu}_e$ fluxes and 
the overall neutrino luminosities are large.   
Additionally, 
by influencing the competition between the charged current neutrino  capture processes 
$\nu_e n  \leftrightarrow  p e^-$ and  $\bar{\nu}_e p  \leftrightarrow  n e^+$~\cite{Qian93}, 
$\nu_e$-$\bar{\nu}_e$ transformation can directly affect the neutron-to-proton ratio, 
which is a key determinant of nucleosynthesis in core-collapse supernova ejecta and compact object ejecta. 

To assess the impact of our findings on neutrino evolution in supernovae,  additional studies are called for.  
First, one needs to have a  full-fledged  (i.e multi-angle) numerical implementation of the coherent QKEs 
in a spherically symmetric model~\cite{shalgar}.  
Moreover, one eventually needs to work out analytically~\cite{collision} and implement numerically 
inelastic collision terms in the QKEs,  including the dependence on the full 
density matrix in flavor and spin space.
\\

\noindent {\bf Acknowledgements} --
This work was supported in part by NSF grant   PHY-1307372  at UCSD,  by  the LDRD Program
at LANL, by the University of California Office of the President,  and by the UC HIPACC  collaboration.
We would also like to acknowledge support from the DOE/LANL Topical Collaboration.
We thank J.~Carlson, J.~F.~Cherry,  and  S.~Reddy  for  useful discussions.
\\

\noindent {\bf Appendix} --  In what follows we outline the steps needed  to obtain the amplitudes 
(\ref{eq:Heff1}-\ref{eq:Heff}) starting from the interaction Lagrangian  (\ref{eq:Lint2}), for the case of 
one massive Dirac neutrino.  
In order to compute  the amplitude (\ref{eq:amplitude}) in perturbation theory,  
we first express  the Lagrangian density  ${\cal L}_{\rm int}$ in terms of free fields, 
written as linear combinations of 
% We express these  expanded in terms of  
creation and annihilation operators for the  helicity states introduced in Sectoin \ref{sect:ensemble}:
\be
\nu (x) =  \int \frac{d^3p}{ (2 \pi)^2  2 E} \ \sum_{h = \pm} \ \left[
a_{\vec{p}, h}      \, u (p,h) \, e^{-i p \cdot x}    
\ + \ 
b_{\vec{p}, h}^\dagger  \, v (p,h) \, e^{i p \cdot x}    
\right]~.  
\ee
Here $p= |\vec{p}|$,  $E = \sqrt{p^2 + m^2}$, and the 
the helicity spinors $u(p,\pm)$ 
%needed to obtain  the neutrino amplitudes  (\ref{eq:Heff1}-\ref{eq:Heff})) 
are given by 
%(the  analogue expression for $v (p, \pm)$ are not needed to obtain  the neutrino amplitudes  (\ref{eq:Heff1}-\ref{eq:Heff})) 
\be
u(p,+) =  \sqrt{E + p} \  \left( 
\begin{array}{c}
r(p)  \, \xi_+ (\hat{p}) \\
\xi_+ (\hat{p})
\end{array}
\right)  
\qquad 
u(p,-) =  \sqrt{E + p} \ 
 \left( 
\begin{array}{c}
 \xi_- (\hat{p}) \\
r(p) \,  \xi_- (\hat{p})
\end{array}
\right)~
\qquad  
r(p) = \frac{m}{E + p}~, 
\ee
with  (denoting  by  $\theta_p, \phi_p$ the polar and azimuthal angles of $\hat{p}$)
\be
\xi_+ (\hat{p}) = 
 \left( 
\begin{array}{c}
 \cos \frac{\theta_p}{2} \\
e^{i \phi_p}  \ \sin \frac{\theta_p}{2} 
\end{array}
\right)~
\qquad 
\xi_- (\hat{p}) = 
 \left( 
\begin{array}{c}
- e^{- i \phi_p}  \ \sin \frac{\theta_p}{2} \\
 \cos \frac{\theta_p}{2} 
\end{array}
\right)
\qquad 
( \vec{\sigma} \cdot \hat{p} )  \ \xi_\pm (\hat{p})  \ = \  \pm  \  \xi_\pm (\hat{p})~. 
\ee

Using the above results and  observing that the interaction 
Lagrangian  density  has the bilinear structure 
${\cal L}_{\rm int} (x)  = \bar{\nu} (x)  \, \Gamma \, \nu (x)$ ($\Gamma$ can be identified from Eq.(\ref{eq:Lint2})),  
we write the interaction Hamiltonian $H_{\rm int}  = - \int d^3x  {\cal L}_{\rm int} (x)$  as 
%
%\be
%H_{\rm int} =  - \sum_{h,h'}  \,  \int \frac{d^3p}{(2 \pi)^3}  \frac{1}{4 E^2}  \  \ 
%a^\dagger_{\vec{p}, h'} \, a_{\vec{p}, h}  \  \  \bar{u} (p,h')  \Gamma u (p,h)            
%+ \dots 
%\ee
%
\be
H_{\rm int} =  - \sum_{h,h'}  \,  \int \frac{d^3p}{(2 \pi)^3}  \frac{1}{4 E^2}  \  \ 
a^\dagger_{\vec{p}, h'} \, a_{\vec{p}, h}  \  \ T_{h' \, h} (p)  + \dots~,  
\qquad \qquad T_{h' \, h} (p)   \equiv   \bar{u} (p,h')  \ \Gamma  \ u (p,h)     ~, 
\ee
where the dots indicate the corresponding  anti-neutrino operators.  
Using this interaction hamiltonian  we compute the amplitudes (\ref{eq:amplitude}) 
in terms of $T_{h' \, h} (p)$.   
The  final results  (\ref{eq:Heff1}-\ref{eq:Heff}) follow after  
explicit calculation of  $T_{h' \, h} (p)$, 
requiring some straightforward Dirac algebra and use of the  relations
%that  can be obtained  by straightforward Dirac algebra and by using the relations
\be
\xi_{\pm}^\dagger  (\hat{p})  \, \vec{\sigma} \,  \xi_\pm (\hat{p})    \ = \ \pm \hat{p} 
\qquad \qquad 
\xi_{\pm}^\dagger  (\hat{p})  \, \vec{\sigma} \,  \xi_\mp (\hat{p})    \ = \ 
e^{\mp i \phi_p} \ \Big( \hat{x}_1 (\hat{p}) \   \mp  \   i \hat{x}_2 (\hat{p})  \Big)~, 
\ee
with $\hat{p}, \hat{x}_{1,2} (\hat{p}) $ defined in (\ref{eq:phat})-(\ref{eq:x2}).

\bibliographystyle{h-physrev}
\bibliography{allref}

\end{document}